\documentclass{aastex61}
\usepackage{xspace}
\usepackage{soul}
\usepackage{color}
\usepackage{graphicx}
\usepackage{natbib}

\usepackage{longtable}

\newcommand\aastex{AAS\TeX}

\shorttitle{\aastex\ Comment on Ravasio Line}

\begin{document}
\AuthorCallLimit=999
\title{Fermi-GBM Team Analysis on the Ravasio Line}

\author[0000-0002-2942-3379]{Eric~Burns}\thanks{Corresponding Author: ericburns@lsu.edu}
\affiliation{Department of Physics \& Astronomy, Louisiana State University, Baton Rouge, LA 70803, USA}

\author[0000-0001-8058-9684]{Stephen~Lesage}
\affiliation{Department of Space Science, University of Alabama in Huntsville, Huntsville, AL 35899, USA}

\author[0000-0002-0587-7042]{Adam~Goldstein}
\affiliation{Science and Technology Institute, Universities Space Research Association, Huntsville, AL 35805, USA}

\author[0000-0003-2105-7711]{Michael~S.~Briggs}
\affiliation{Department of Space Science, University of Alabama in Huntsville, Huntsville, AL 35899, USA}
\affiliation{Center for Space Plasma and Aeronomic Research, University of
Alabama in Huntsville, Huntsville, AL 35899, USA}

\author[0000-0002-2149-9846]{Peter~Veres}
\affiliation{Department of Space Science, University of Alabama in Huntsville, Huntsville, AL 35899, USA}
\affiliation{Center for Space Plasma and Aeronomic Research, University of
Alabama in Huntsville, Huntsville, AL 35899, USA}

\author[0000-0002-6657-9022]{Suman Bala}
\affiliation{Science and Technology Institute, Universities Space Research Association, Huntsville, AL 35805, USA}

\author[0009-0003-7797-1243]{Cuan~de Barra}
\affiliation{School of Physics, Centre for Space Research, Science Center North, University College Dublin, Dublin 4, Ireland}

\author[0000-0001-9935-8106]{Elisabetta Bissaldi}
\affiliation{Dipartimento Interateneo di Fisica dell'Università e del Politecnico di Bari, 70125 Bari, Italy}
\affiliation{Istituto Nazionale di Fisica Nucleare, Sezione di Bari, 70125 Bari, Italy}

\author[0009-0003-3480-8251]{William~H~Cleveland}
\affiliation{Science and Technology Institute, Universities Space Research Association, Huntsville, AL 35805, USA}

\author{Misty~M~Giles}
\affiliation{Jacobs Space Exploration Group, Huntsville, AL 35806, USA}

\author[0000-0002-2926-0469]{Matthew Godwin}
\affiliation{Department of Space Science, University of Alabama in Huntsville, Huntsville, AL 35899, USA}

\author[0000-0001-9556-7576]{Boyan~A.~Hristov}
\affiliation{Center for Space Plasma and Aeronomic Research, The University of Alabama in Huntsville, Huntsville, AL 35899}

\author[0000-0002-0468-6025]{C.~Michelle~Hui}
\affiliation{Astrophysics Branch, NASA Marshall Space Flight Center, Huntsville, AL 35812, USA}

\author[0000-0001-9201-4706]{Daniel~Kocevski}
\affiliation{Astrophysics Branch, NASA Marshall Space Flight Center, Huntsville, AL 35812, USA}

\author[0000-0002-2531-3703]{Bagrat~Mailyan}
\affiliation{Department of Aerospace, Physics and Space Sciences, Florida Institute of Technology, Melbourne, FL 32901, USA }

\author[0000-0002-0380-0041]{Christian~Malacaria}
\affiliation{INAF-Osservatorio Astronomico di Roma, Via Frascati 33, I-00076, Monte Porzio Catone (RM), Italy}

\author[0000-0002-1477-618X]{Sheila~McBreen}
\affiliation{School of Physics, Centre for Space Research, Science Center North, University College Dublin, Dublin 4, Ireland}

\author[0000-0003-1626-7335]{Robert~Preece}
\affiliation{Department of Space Science, University of Alabama in Huntsville, Huntsville, AL 35899, USA}

\author[0000-0002-7150-9061]{Oliver~J.~Roberts}
\affiliation{Science and Technology Institute, Universities Space Research Association, Huntsville, AL 35805, USA}

\author[0000-0002-0602-0235]{Lorenzo~Scotton}
\affiliation{Department of Space Science, University of Alabama in Huntsville, Huntsville, AL 35899, USA}
\affiliation{Center for Space Plasma and Aeronomic Research, University of Alabama in Huntsville, Huntsville, AL 35899, USA}
\author[0000-0002-0221-5916]{A.~von Kienlin}
\affil{Max-Planck-Institut f\"{u}r extraterrestrische Physik, Giessenbachstrasse 1, D-85748 Garching, Germany}

\author[0000-0002-8585-0084]{Colleen~A.~Wilson-Hodge}
\affiliation{Astrophysics Branch, NASA Marshall Space Flight Center, Huntsville, AL 35812, USA}

\author[0000-0001-9012-2463]{Joshua~Wood}
\affiliation{Astrophysics Branch, NASA Marshall Space Flight Center, Huntsville, AL 35812, USA}

\begin{abstract}
The prompt spectra of gamma-ray bursts are known to follow broadband continuum behavior over decades in energy. GRB~221009A, given the moniker the brightest of all time (BOAT), is the brightest gamma-ray burst identified in half a century of observations, and was first identified by the Fermi Gamma-ray Burst Monitor (GBM). On behalf of the Fermi-GBM Team, \citet{lesage2023fermi} described the initial GBM analysis. \citet{ravasio2024mega} report the identification of a spectral line in part of the prompt emission of this burst, which they describe as evolving over 80~s from $\sim$12~MeV to 6~MeV. We report a GBM Team analysis on the Ravasio Line: 1) We cannot identify an instrumental effect that could have produced this signal, and 2) our method of calculating the statistical significance of the line shows it easily exceeds the 5$\sigma$ discovery threshold. We additionally comment on the claim of the line beginning at earlier time intervals, up to 37~MeV, as reported in \citet{zhang2024observation}. We find that it is reasonable to utilize these measurements for characterization of the line evolution, with caution. We encourage theoretical studies exploring this newly discovered gamma-ray burst spectral feature, unless any rigorous alternative explanation unrelated to the emission from GRB 221009A is identified.
\end{abstract}

\section{Context} \label{sec:intro} 
This paper serves as a comment on published results and includes information only relevant for the specific analysis contained within. Context and background on GRB\,221009A as the brightest gamma-ray burst is well described in the literature. The Fermi-GBM Team analysis is reported in \citet{lesage2023fermi}. The GBM trigger time T$_0$ is 2022 October 9 at 13:16:59.99 UTC. All times here are referenced against this time. The GBM data has significant issues caused by two separate instrumental limitations covering the intervals from T0+219~s to T0+277~s, as well as T0+508~s to T0+514~s.

\citet{ravasio2024mega} identified a signal interpreted as an MeV line. They identify the line first at $\sim$14\,MeV beginning at T0+280~s and consistently evolving down to $\sim$6\,MeV out to T0+320~s.\footnote{As this line is not at a singular energy it is not precisely described by a numerical value. As there are atomic and molecular lines observed in the follow-up of this event, it is not appropriate to call this the BOAT line. Thus, for brevity, we refer to this as the Ravasio Line, following the established tradition in our field of naming things after Italians.} As summarized in the introduction of \citet{ravasio2024mega}, claims of GRB lines in the prompt spectrum are controversial, being connected to the Great Debate on the cosmological or Galactic origin of GRBs \citep{Nemiroff95greatdebategrb}. The claimed signals either arose from instrumental effects or from improper statistical handling. Prior claims of lines in prompt GRB spectra were ultimately disproven through dedicated studies by the instrument teams on the data from the Burst And Transient Source Experiment (BATSE) on-board the Compton Gamma-Ray Observatory \citep[e.g.][]{1994ApJ...433L..77P, 1994ApJ...434..560B, 1995ApJ...447..289B, 1996ApJ...458..746B, 1996AIPC..384..153B, 1997ApJ...485..747B, 1999ApL&C..39..237B}.

We thus write this comment for two reasons. First, as the instrument team, we have a deep understanding of the underlying instrument and possible instrumental effects. Second, we utilize our preferred empirical method of statistical preference for a more complicated model over another to quantify significance.

\section{Comment on Ravasio et al. 2024}
The most direct method of confirming the claim in \citet{ravasio2024mega} would be identification in a separate instrument. This would very strongly argue against an instrumental effect, or require one affecting both instruments. It would also provide additional statistical evidence. Unfortunately, as described in \citet{ravasio2024mega}, despite dozens of observations of the prompt phase, no other instrument has sufficient data at these energies. We are thus limited to working with the Fermi-GBM data.

\subsection{Possibility of Instrumental Effects}
The analysis of potential instrumental effects presented by \citet{ravasio2024mega} is accurate. In support of their conclusion, we provide the following arguments to further substantiate that no known instrumental artifacts could account for the observed behavior of the spectral line described in \citet{ravasio2024mega}.

Satellites are constantly bombarded by high energy particles, some of which interact and cause the spacecraft itself to be radioactive, and some which annihilate directly. The most prominent are the 511~keV positron annihilation line and the 2.2~MeV neutron capture line due to the hydrazine propellant \citep{Meegan2009}. Indeed given the spectral resolution of Bismuth Germanate (BGO) detectors on-board GBM at these energies, this may be the origin of the 2.1\,MeV line identified in GRB\,221009A \citep{jiang2024discovery}. There are also nuclear lines such as the 1.46~MeV line from $^{40}$K. In principle, such lines can be caused by photonuclear reactions as gamma-rays interact with either the Fermi spacecraft or the atmosphere. They have been seen from Terrestrial Gamma-ray Flashes \citep{bowers2017gamma,rutjes2017tgf}; as these events have photons with similar energies to those seen in the BOAT, these interactions may also be at play here. However, the cross sections are too small to explain the reported line flux. Regardless of the physical interaction, such lines are all emitted at a fixed energies, which is contrary to the monotonic and continuous evolution identified in \citet{ravasio2024mega}.

The GBM Team paper for GRB\,221009A includes discussions of limitations imposed by two distinct instrumental effects. GBM data types include photon-by-photon time-tagged event (TTE) data  and binned data referred to as CSPEC and CTIME. When the rate of events observed by the sum of all detectors exceeds 375,000 counts per second TTE data is lost \citep{Meegan2009}. \citet{ravasio2024mega} uses CSPEC data, so this effect is irrelevant.

The second effect is pulse-pileup which can cause spectral distortion and should thus be carefully considered when investigating a claim of a new spectral feature. When a given event hits a detector, an electronic pulse is generated whose height is related to the inferred energy. Each pulse has a fixed processing deadtime of 2.6 microseconds, except for very large pulses in the highest channel, for which a deadtime of 10 microseconds is applied \citep{Meegan09}. The most common form of pulse pileup is when multiple events arrive within a single pulse window resulting in an incorrect inferred pulse height and two or more events will be recorded as one. This causes two related problems: fewer events will be recorded than actually interacted in the detector, and the recorded energy channel will be incorrect (it can either be lower or higher than the height from the original photon) causing spectral distortion. These effects in GBM are studied in detail in \citet{chaplin2013analytical} and \citet{bhat2014fermi}. We return to description of this effect in the next section. 

The detectors utilized in \citet{ravasio2024mega} have the best viewing angle to the source, corresponding to the highest count rate detectors. They exit the pulse-pileup region before T0+280~s, as stated in our original public release of the Bad Time Intervals\footnote{\url{https://fermi.gsfc.nasa.gov/ssc/data/analysis/grb221009a.html}}. Additionally, the instrument recovery times are sub-second. Thus, the time intervals considered in \citet{ravasio2024mega} are not affected by pulse-pileup and this instrumental effect is also not relevant here.

Another possible consideration is afterglow in the scintillators themselves (note this a detector effect, and not referring to the GRB afterglow arising from synchrotron as the jet interacts with the surrounding medium). This is scintillation light which remains after the initial event interaction in the detector. Given the exceedingly high rates there may be afterglow present. BGO has short and minimal afterglow, with representative values for BGO crystals being afterglow at the 0.015\% level after 20~ms\footnote{\url{https://luxiumsolutions.com/radiation-detection-scintillators/crystal-scintillators/bgo-bismuth-germanate}}. We find this to be an unlikely explanation. First, the primary afterglow timescale is much faster than a second, so the count rate during the saturated intervals are unlikely to effect the intervals of interest for the line. This is strengthened by the lack of identification in bursts with higher peak rates than the line intervals for the BOAT \citep{ravasio2024mega}. Additionally, the afterglow percentage is $\sim$0.015\%. As the inferred photon flux above 250\,keV of the line in the 300-320~s interval is 1.9\%, being orders of magnitude too large. Lastly, the afterglow should approximately match the originating spectrum, whereas the line is inconsistent with the dominant component.

We are unaware of any instrumental effect which may produce a line that appears at one energy and moves to another. We encourage any community member who is aware of a possibly relevant effect to submit a paper for publication in an appropriate journal. Until such a paper is published, we conclude that an instrumental origin for the line is either excluded or requires some effect unknown to the Fermi-GBM Team.

\subsection{Statistical Significance}
The next key question is the statistical significance of the signal. This is not trivial to assess. GBM (and similar detectors) have non-invertible detector responses. This requires forward-folding of a model through the detector response and then comparison with data. This then makes the inferred flux (density) of a given component inherently model dependent, and statistical significance calculations complicated. 

Utilizing likelihood functions, statistical significance can be determined by comparing a likelihood for a given spectral functional form against the likelihood for a second spectral form. In this case, comparing the likelihood for the continuum-only model (null hypothesis) against that from the continuum+line model (alternative hypothesis). The models considered in this manuscript are typical continuum models utilized in the study of GRB spectra and a Gaussian line. 

Typically a likelihood ratio test would provide a statistical measure for accepting or rejecting the alternative hypothesis, but this particular case violates some regularity conditions required for the likelihood ratio test\citep[e.g.][]{protassov2002statistics}. \citet{algeri2019searching} describe regularity conditions for approaches which rely on Wilks' theorem. These include being \textit{identifiable}, where different parameter values specify distinct models. In our case, the amplitude of the Gaussian line is zero in the null hypothesis case. When combined with the lack of {\it a priori} knowledge on the centroid of the line, this situation is not identifiable. The lack of knowledge is related to the Look Elsewhere Effect since we do not have a prior theoretical expectation on where this line should occur and identification of a line anywhere in the GBM energy range may be reported as interesting, there is an effective trials factor due to the range of interest. 

A second regularity condition, referred to as \textit{interior} in \citet{algeri2019searching}, requires that the parameter measures of interest do not reside on the bounds considered for that parameter. In our case, the amplitude of the Gaussian line component is required to be greater than or equal to zero, and thus this boundary regularity condition is also not met.

We note that the regularity conditions being violated does not necessarily mean that the likelihood ratio test will fail to provide an approximate determination of significance, but it is not guaranteed to be accurate, and therefore requires verification, which can be done through empirical determination of the p-value through simulation. The difference between the log-likelihood of the null hypothesis and the alternative hypothesis can be compared to a background distribution generated by simulations, where the p-value is the fraction of simulations in excess of the critical delta log-likelihood value. However, this is computationally expensive. The need to quantify line significance in astronomy is closely related to seeking resonances in high-energy particle physics. As such, numerous techniques from this field have been developed to handle violations of the regularity conditions, minimize computation, or provide approximate solutions and are often applied in our field. 

The Akaike Information Criterion (AIC) \cite{akaike1974new} is utilized in the main portion of \citet{ravasio2024mega}; it is based on information theory to determine if a preference exists between two models (avoiding any statement on absolute goodness-of-fit). The AIC avoids some regularity conditions (e.g., it does not require nested models), but it does have others (e.g. having large enough counts to be asymptotic). The AIC is affected by the boundary problem \citep{mitchell2022generalized}. This means it requires additional verification to understand the validity of the its statistical significance assignment. 

\citet{ravasio2024mega} utilize the technique described in \citet{gross2010trial}. This is a method to extrapolate from a small number of simulations while accounting for the Look Elsewhere Effect and the non-identifiability problem. This method still suffers from the boundary problem (though see \citet{chernoff1954distribution}); however, this method is still an approximation, and, when valid, may be prone to being overly conservative \citep{algeri2019searching}. Again, this approach may be approximately valid based on the problem it is applied to, but it is not statistically guaranteed to be valid.

Thus, we seek a purely empirical determination of the statistical significance of the line. \citet{ravasio2024mega} find the line to be significant in their Interval 5 (280 to 300~s) and Interval 6 (300 to 320~s), with possible extension into Interval 7 (320 to 340~s) and Interval 8 (340 to 360~s). They additionally sub-divide Interval 5 into 4 intervals, 280-285, 285-290, 290-295, and 295-300~s, and Interval 6 into two intervals, 300-310 and 310-320~s. They find a preference for the continuum to be a Smoothly Broken Power Law (SBPL) \citep{gruber2014fermi} in the intervals 280-300, 280-285, 285-290, 290-295, 295-300, 300-320, and 300-310. For intervals 310-320, 320-340, and 340-360 they find a preference for the continuum to contain the SBPL as well as an additional power-law (PL). These form their continuum models in each interval, which is the null hypothesis. In each case the alternative hypothesis is the continuum model and a Gaussian line. 

Because this significance is being calculated \textit{ex post facto} we will minimize our bias by using standard GBM selections and match decisions made for \citet{lesage2023fermi}, which were made before the team was aware of the line feature. In brief, we utilize the same background fitting intervals and polynomial order, we use data from GBM Sodium Iodide (NaI) detectors from 45\,keV (due to structured residuals at low energies seen in this event) to 900\,keV and data from BGO from 400\,keV (due to unmodeled spacecraft blockages along the path to the source) to 39\,MeV. In intervals with a continuum best fit with a SBPL, we fix the break scale to be 0.3, matching our standard catalog practice \citep{kaneko2006complete,gruber2014fermi,Poolakkil+21GBMspeccat}. We do not remove the iodine k-edge, as we are not concerned with structure around this energy ($\sim$33~keV). We do not allow for an effective area correction between detectors.

We do deviate from standard practice and past decisions in two cases. First, the SBPL break scale is free in the intervals with continuum best fit by SBPL+PL, as it is unclear if the extra component would result in the same preferred break scale. Second, in \citet{lesage2023fermi} we utilize detectors NaI~4, NaI~8, and BGO~1. We add NaI~6 as it has a good viewing angle through the timescales considered. We additionally include BGO~0, as this should add sensitivity in a search for lines at the energy of interest. \citet{ravasio2024mega} utilize similar detector selections.

Our procedure for each interval is as follows. We fit the data using the Gamma-ray Data Tools (GDT)\footnote{\url{https://github.com/USRA-STI/gdt-core}} for the specified continuum model from \citet{ravasio2024mega} and for the continuum model with a Gaussian line. The difference in the Pstat value (likelihood for a Poisson signal and a background model with negligible model variance\footnote{Pstat is the special case of PGstat (Poisson data and Gaussian background) where the background model variance approaches zero, such as in the case of a long background exposure. Pstat is often preferable in this case because PGstat is more expensive to calculate, and some solvers may fail in certain scenarios when using PGstat}; this is equivalent to the c-stat fit statistic from {\tt RMfit}, adopted as GBM standard practice) is our test statistic. We generate spectral simulations by assuming the null hypothesis is true, (i.e. continuum-only is the true spectrum), convolving this spectrum with our response, adding the background counts and Poisson fluctuations, and fit the null and alternative hypotheses by finding the maximum likelihood ($-$Pstat). The difference in the Pstat values between the two hypotheses are then used to calculate the $\Delta$Pstat value. The p-value is then simply the number of events in the simulations greater than the critical value found by fitting the real data and divided by the number of simulations. In cases where no simulations exceed the measured value we set the p-value as $<$1 divided by the number of simulations.

We use the Nelder-Mead minimization algorithm \citep{gao2012implementing} as implemented in SciPy \citep{virtanen2020scipy}. Nelder-Mead provides robust parameter value and error measurements, but is susceptible to becoming trapped in a local minimum. To counter this, we seed parameter values from \citet{ravasio2024mega}. As an additional check, we fit each interval with the classic GBM spectral analysis software, RMfit\footnote{\url{https://fermi.gsfc.nasa.gov/ssc/data/analysis/gbm/}}, finding reasonable agreement in the values reported in \citet{ravasio2024mega}.

The fits from the GDT implementation on the measured data are largely consistent, within errors, with those reported in \citet{ravasio2024mega}. In the 310-320~s interval we fail to properly measure the high-energy index of the SBPL, though this is not a problem when using only BGO 1 (i.e., excluding BGO 0). The only trend of disagreement is our fits tend to measure a wider Gaussian line width than \citet{ravasio2024mega}. When we fit with only BGO 1 (excluding BGO 0) the inferred width narrows (for intervals where the fit succeeds), which explains most of the difference between our values and those in \citet{ravasio2024mega}. A GBM BGO flight spare was calibrated on the ground to 17.6~MeV \citep{bissaldi2009ground}, while the BGO on the spacecraft have in-flight validation up to 6~MeV \citep{ackermann2012fermi}. This may suggest that the wider inferred width when using both BGOs is not due to inter-calibration issues, but this is not certain.

\begin{figure}
\centering
\includegraphics[scale=0.7]{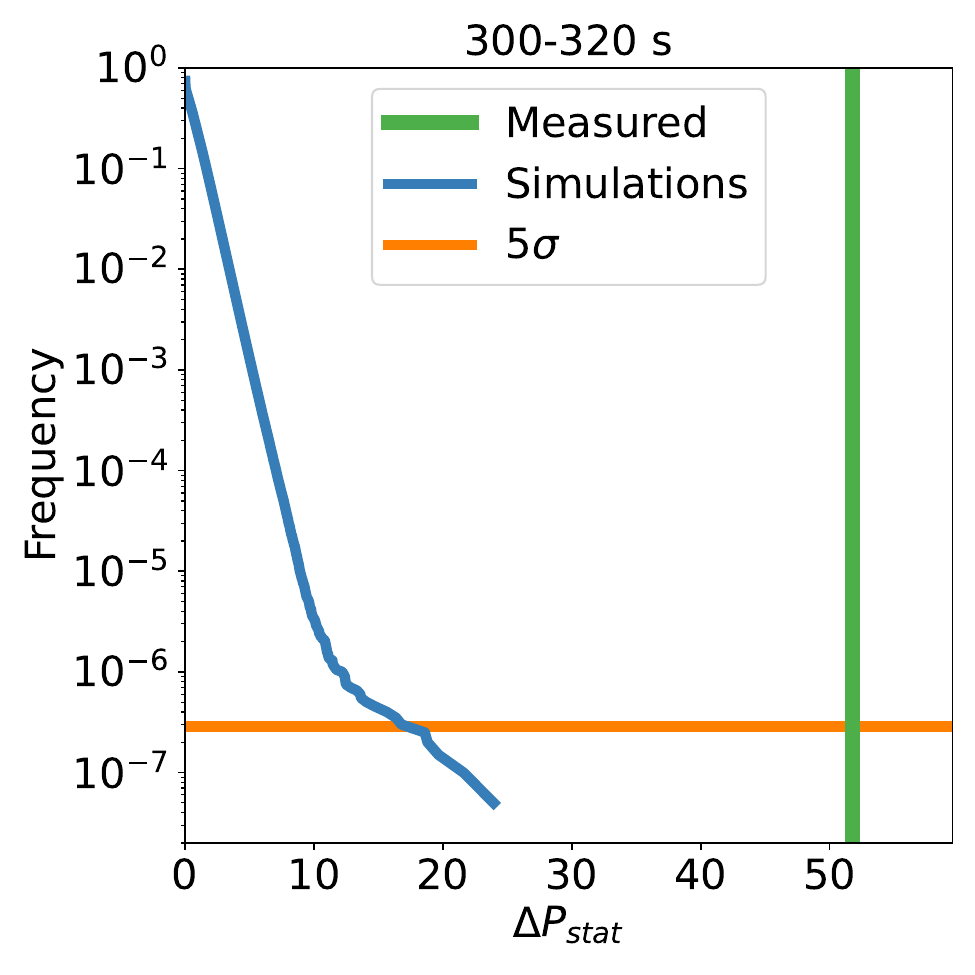}
\caption{The comparison between the measured $\Delta$Pstat and the values from simulations. We performed 20 million simulations, resulting in a p-value constraint of $\sim$5E-8. Despite the large number of simulations, no $\Delta$Pstat is within a factor of two of the measured value. Thus, this interval alone clearly exceeds the standard 5$\sigma$ discovery significance.}
\label{fig:significance}
\end{figure}

\begin{figure}
    \centering
\includegraphics[width=0.32\textwidth]{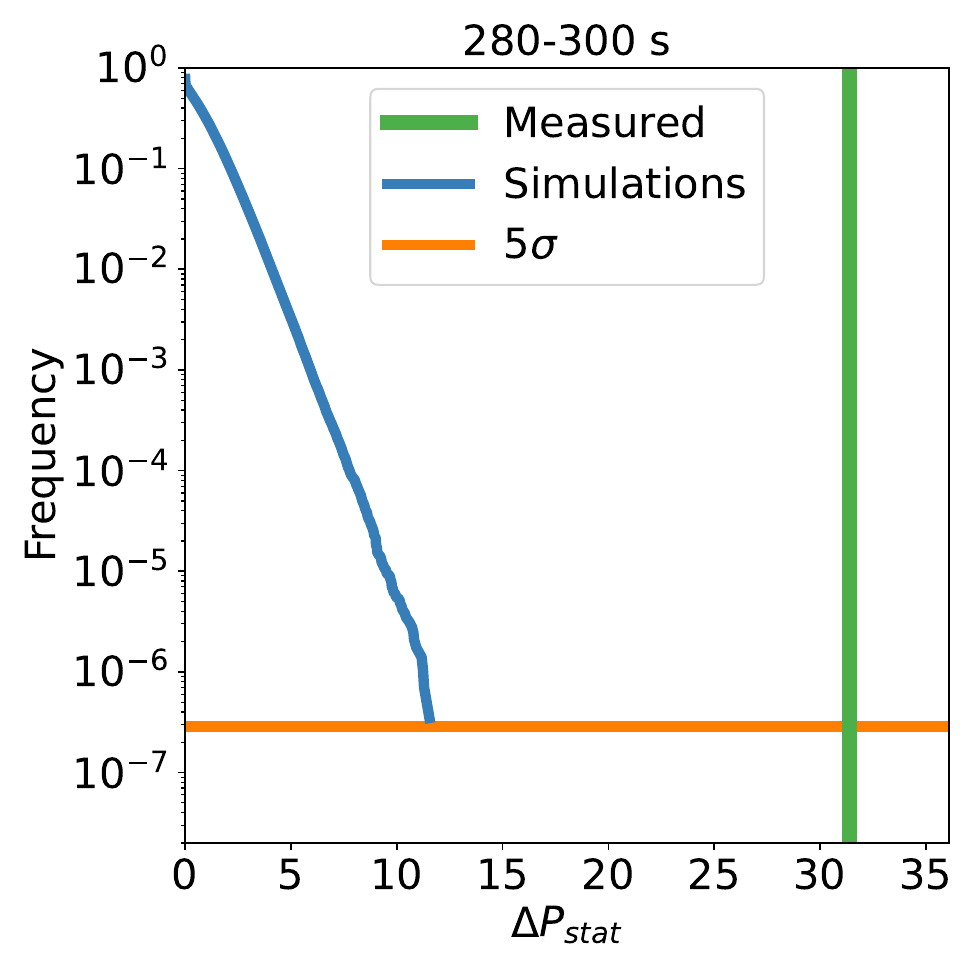}
\includegraphics[width=0.32\textwidth]{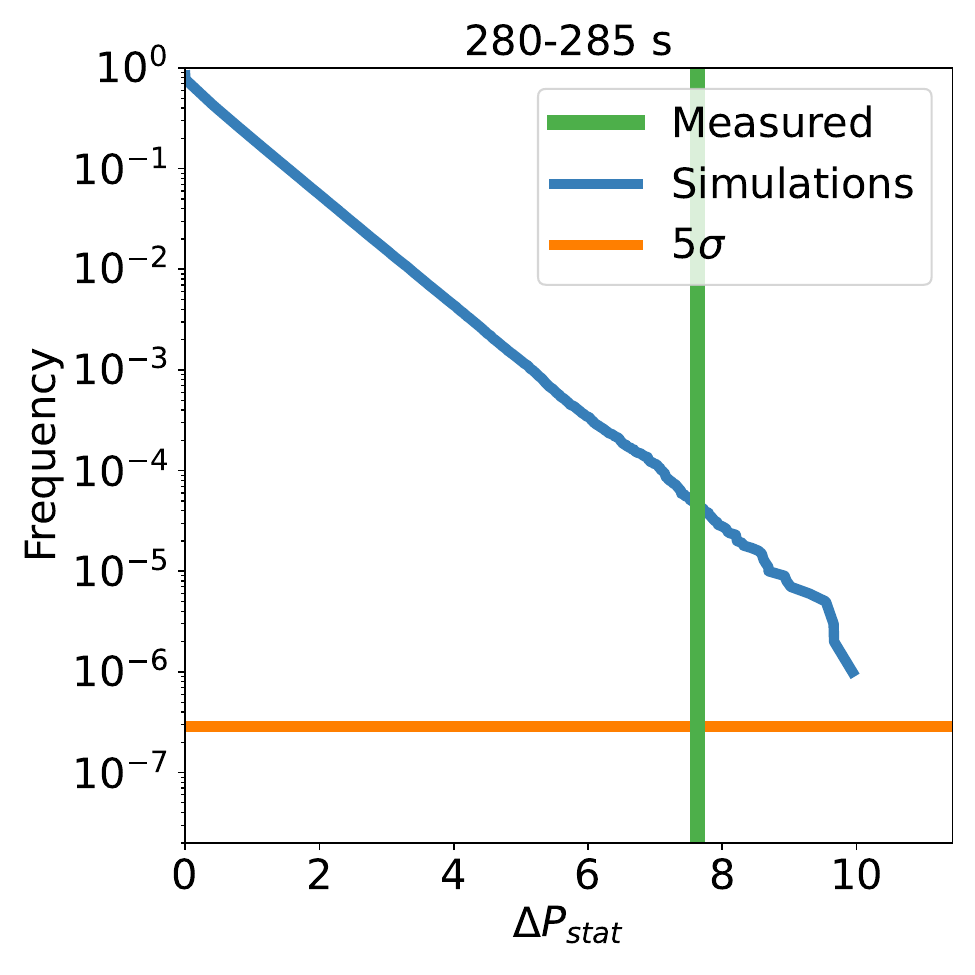} 
\includegraphics[width=0.32\textwidth]{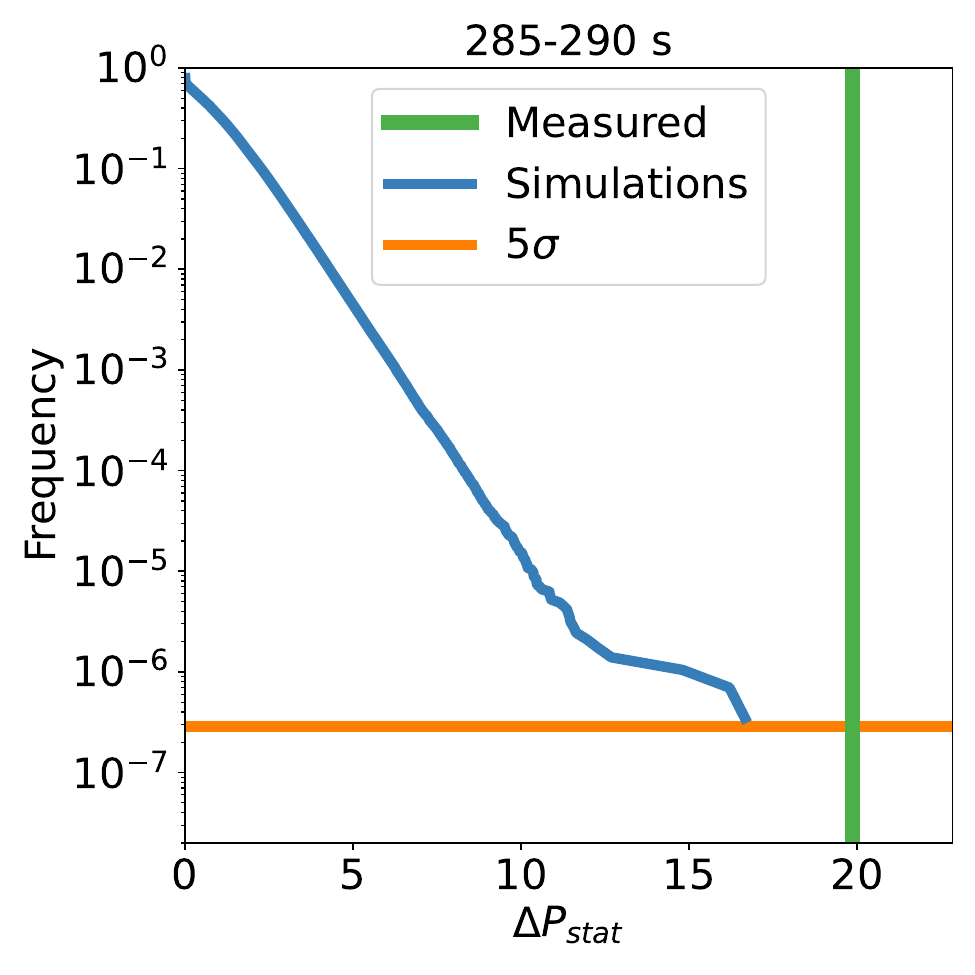}
\includegraphics[width=0.32\textwidth]{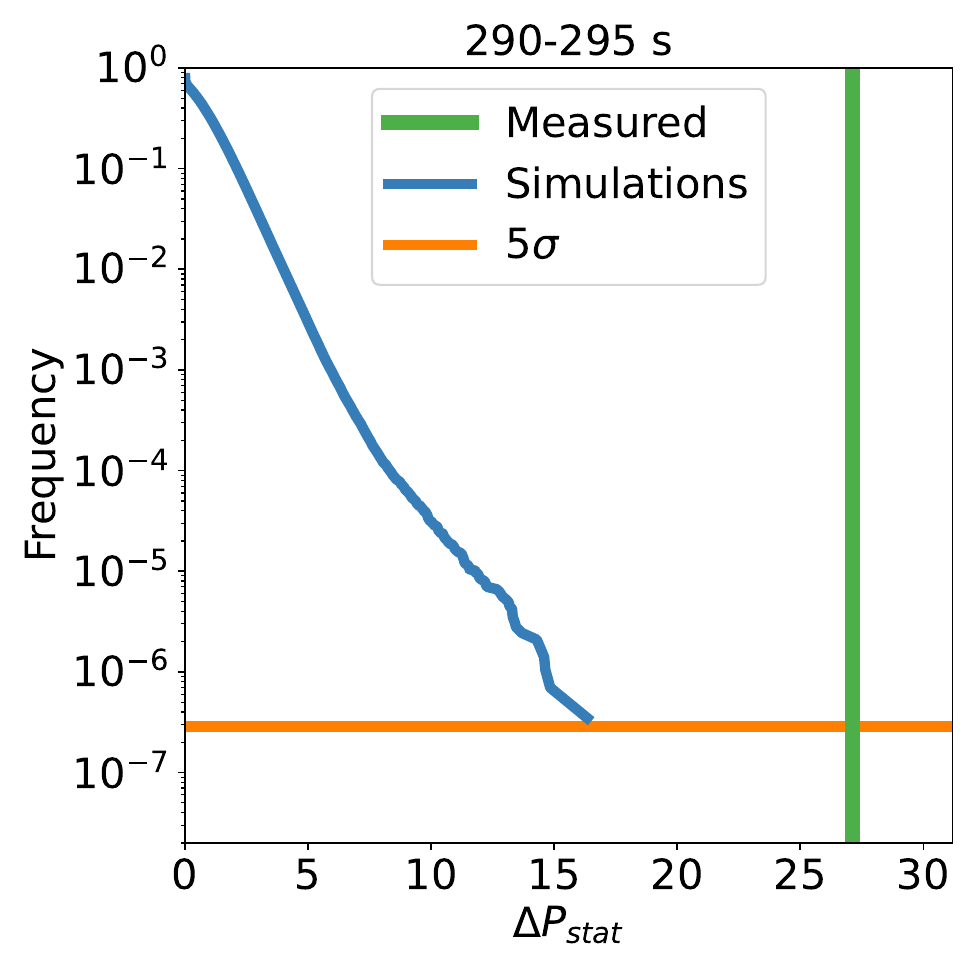}
\includegraphics[width=0.32\textwidth]{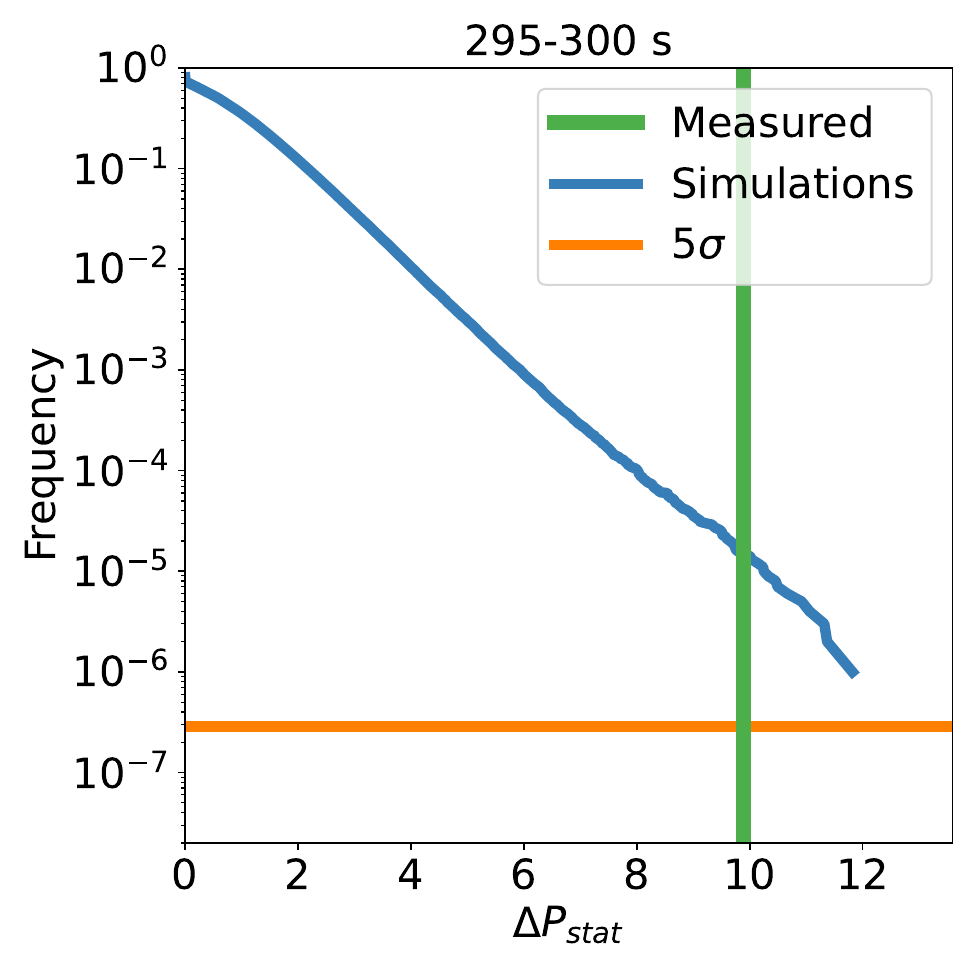}
\includegraphics[width=0.32\textwidth]{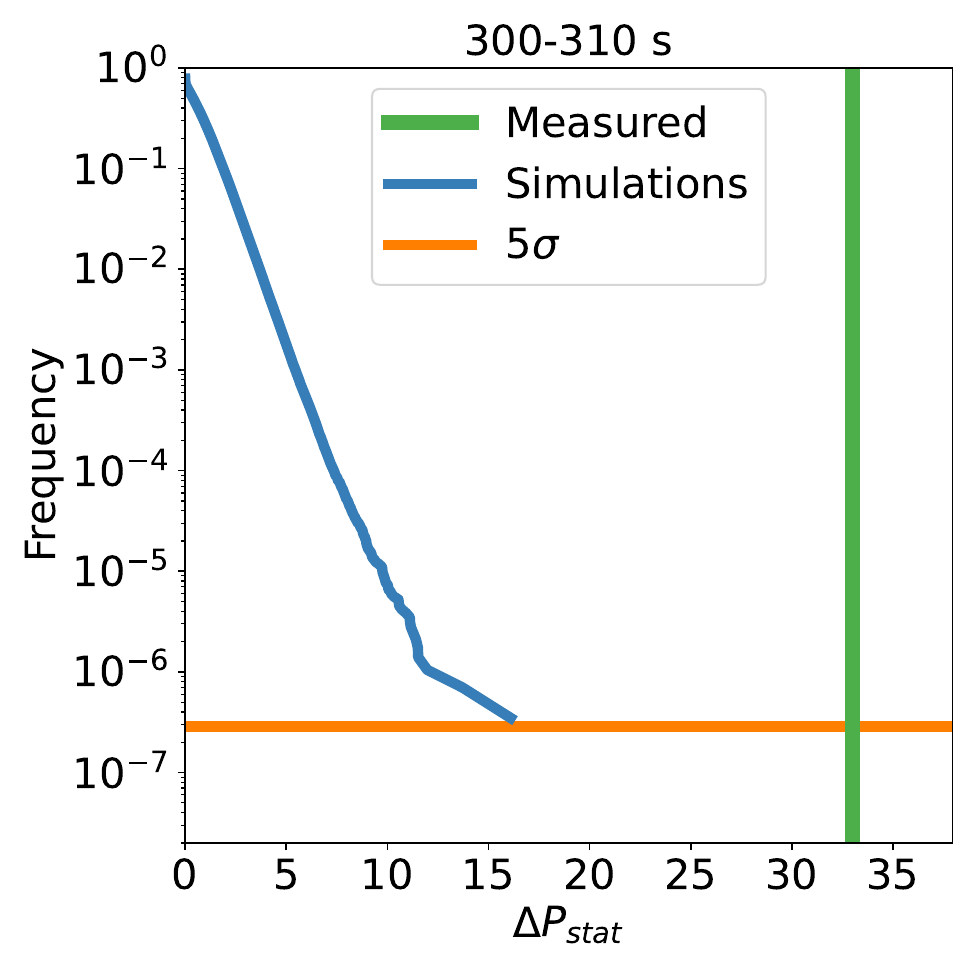}
\includegraphics[width=0.32\textwidth]{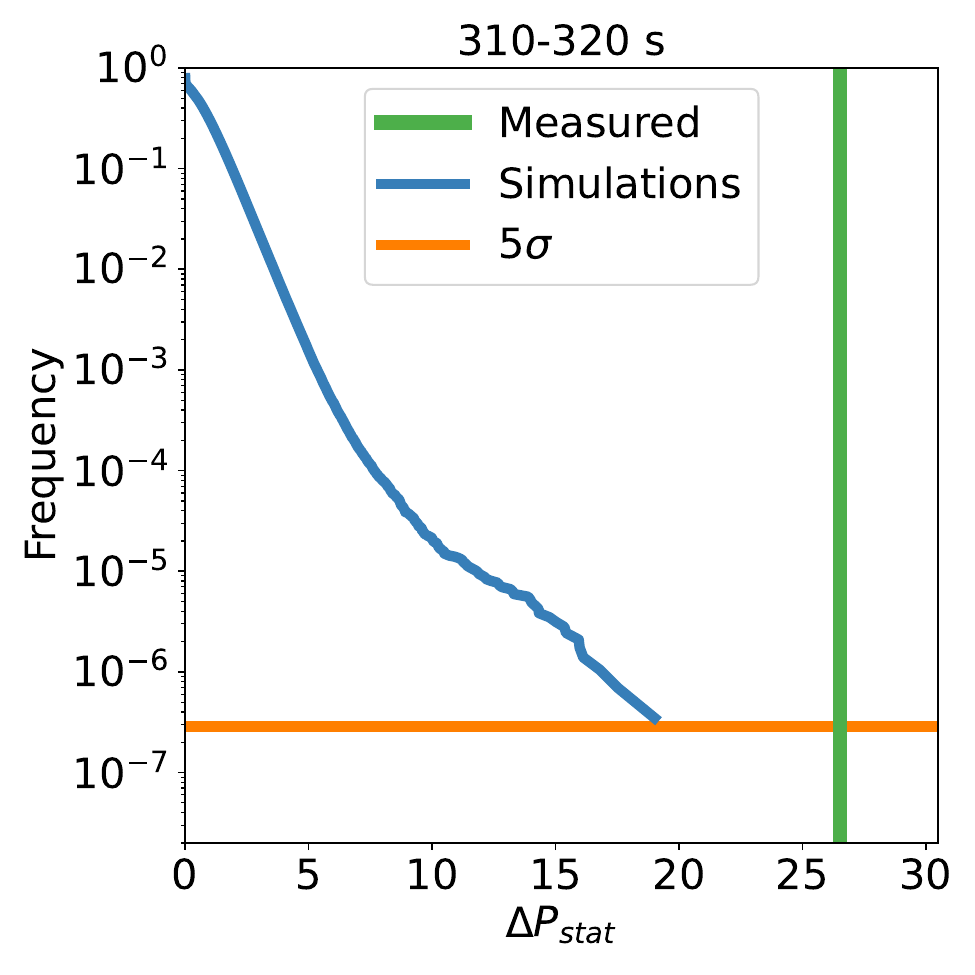}
\includegraphics[width=0.32\textwidth]{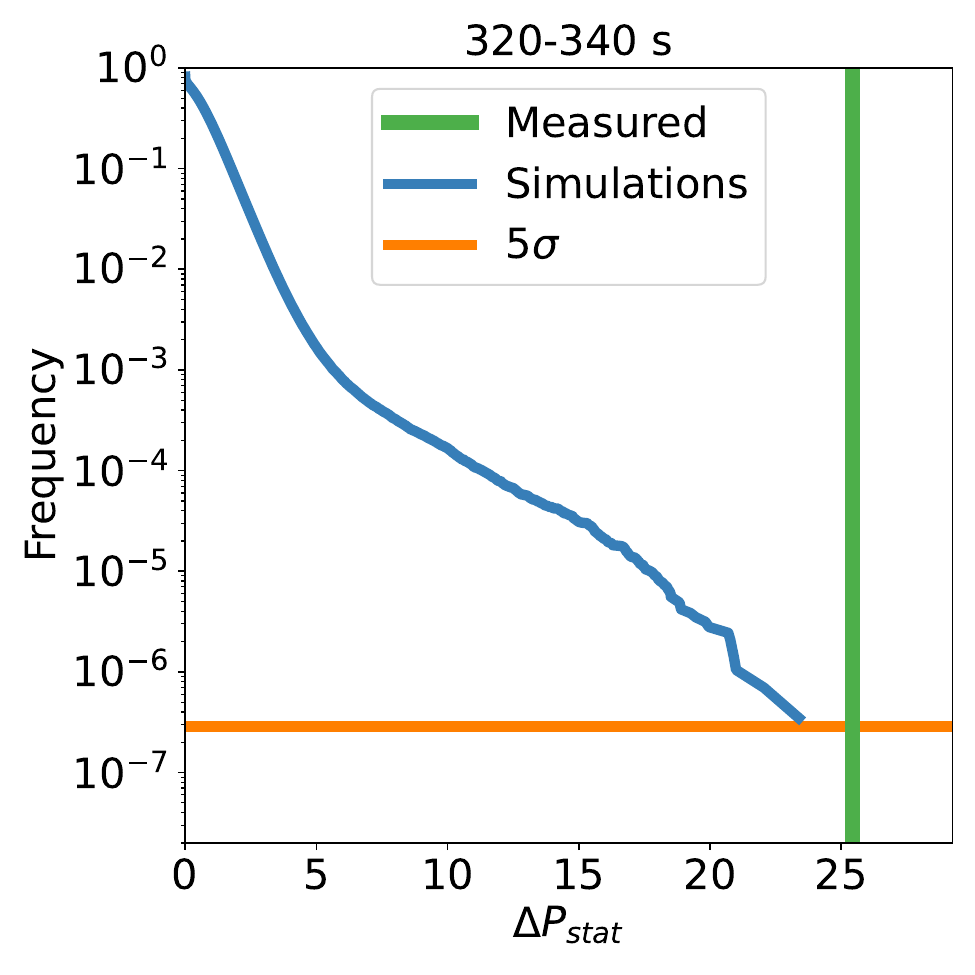}
\includegraphics[width=0.32\textwidth]{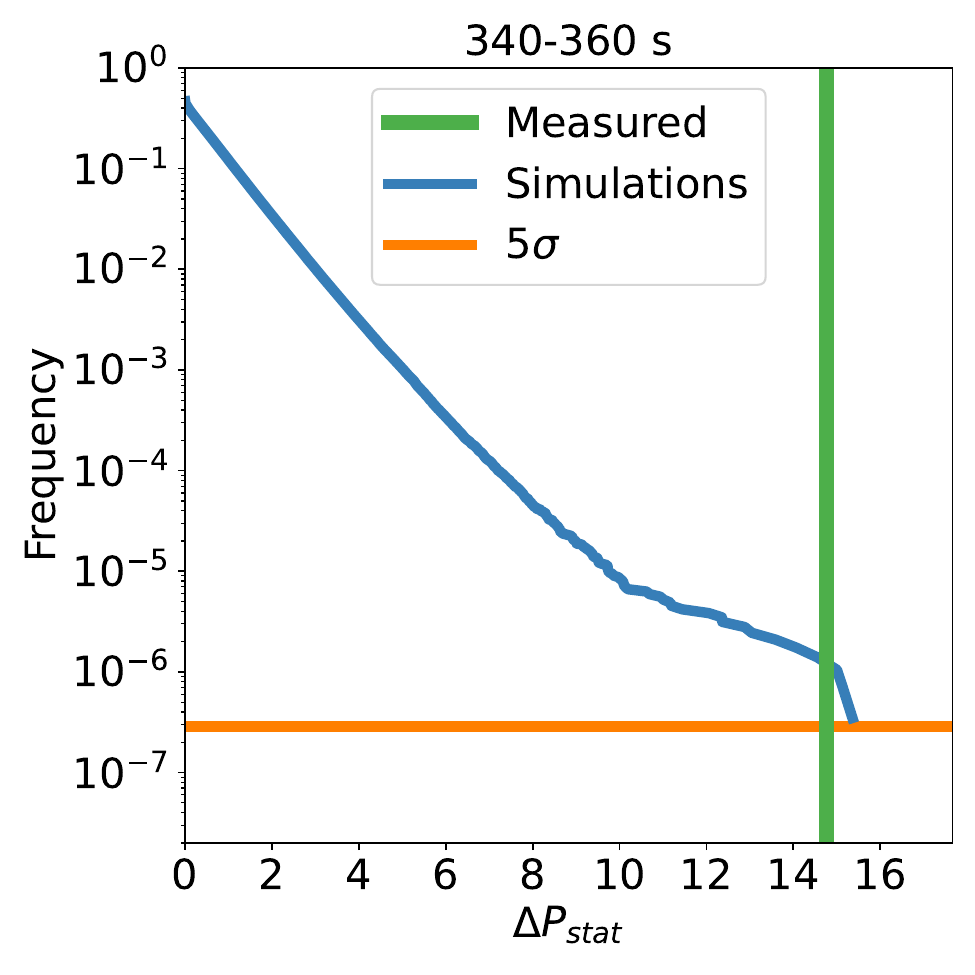}
    \caption{The same as Figure~\ref{fig:significance}, but for the other intervals of interest.}
    \label{fig:other_significances}
\end{figure}

The comparison of the measured $\Delta$Pstat against the simulations are shown in Figures~\ref{fig:significance} and \ref{fig:other_significances}. Quantitative information on statistical significance for each interval is summarized in Table\ref{tab:results}. We find that Interval 6, 300-320~s, is well in excess of 5$\sigma$ discovery significance. Several additional intervals also exceed this value. The line is, without a doubt, statistically significant.

\begin{table}
\centering
\begin{tabular}{|c|c|c|c|c|c|}
\hline
Interval & $\Delta$Pstat & p-value & sigma \\
\hline
280-300 & 31.4 & $<$3.5e-07 & $>$5.0\\
280-285 & 7.6 & ~4.5e-05 & 3.9\\
285-290 & 19.9 & $<$3.5e-07 & $>$5.0\\
290-295 & 27.1 & $<$3.5e-07 & $>$5.0\\
295-300 & 9.9 & 1.5e-05 & 4.2\\
300-320 & 51.8 & $<$5.0e-08 & $>$5.3\\
300-310 & 33.0 & $<$3.5e-07 & $>$5.0\\
310-320 & 26.5 & $<$3.5e-07 & $>$5.0\\
320-340 & 25.4 & $\sim$3.5e-07 & $\sim$5.0\\
340-360 & 14.8 & 1.0e-06 & 4.7\\
\hline
\end{tabular}
\caption{Our measured statistical significance of the intervals analyzed in \citet{ravasio2024mega}. According to \citet{ravasio2024mega} the 300-320~s interval is the most significant. As such, we ran 20,000,000 simulations for the 300-320~s interval to unambiguously determine if the significance exceeds 5$\sigma$, finding that it does. 280-285~s and 295-300~s are run for 1,000,000 simulations, which was sufficient to measure significance. For these intervals the p-value is measured directly as the fraction of simulations whose $\Delta$Pstat exceeds the measured value. The other intervals are run for 2,867,000 iterations, allowing for a nominal test of 5$\sigma$ discovery. For cases where no simulated values approach the measured value we take 1/$N_{sims}$ as a lower limit on the p-value; if the simulated values approach the measured one, we take 1/$N_{sims}$ as the approximate value.}\label{tab:results}
\end{table}

We note that this method is testing only the statistical significance of the line assuming the continuum model is the true spectrum, which may not be true. Applying the same procedure to some intervals but fitting a Band \citep{1993ApJ...413..281B} function gives similar results.

\section{Comment on Zhang et al. 2024}
Following the discovery of the Ravasio Line in \citet{ravasio2024mega}, \citet{zhang2024observation} performed additional investigations to characterize the line by combining data at lower energies from GECAM-C with data covering the line energies from the Fermi-GBM BGO. Their analysis includes the same intervals as \citet{ravasio2024mega} with a few non-trivial differences. First, they use GECAM-C data instead of Fermi-GBM NaI data. Second, they use a likelihood that considers Poisson background and Poisson signal, instead of the likelihood fit statistic in \citet{ravasio2024mega} which considers Gaussian background and Poisson signal. Third, \citet{zhang2024observation} do not consider the same spectral forms, resulting in selection of a best-fit spectrum of the Band function \citep{1993ApJ...413..281B} rather than the SBPL (and related) in \citet{ravasio2024mega}. Despite these differences, the trends of the fit parameters and the inclusion of an extra power-law at later times agree. \citet{zhang2024observation} find several of these intervals to be statistically significant, adding further support to the discovery claims in \citet{ravasio2024mega}. Specific differences in statistical significance are likely due to the different choices mentioned above.

\citet{zhang2024observation} analyze additional intervals during the brightest portion of the burst, claiming additional evidence for the line at intervals 246-256~s, 270-275~s, and 275-280~s. The GBM data used here is CSPEC, avoiding global saturation effects of the GBM TTE data. Thus, they must only determine if the individual GBM BGO detectors are experiencing pulse pileup in their intervals of interest. In principle, this is not trivial, as the measured rate when pulse pileup is occurring is lower than the true rate. For the bright intervals considered in \citet{zhang2024observation} the GBM NaI are experiencing pulse pileup, but the BGO fall below our guideline rate of 60,000 counts per second. Thus, the BGO data is likely to be valid in these intervals; however, because of the data issues which occur during pulse pileup we cannot be certain that intervals with count rates below our guideline threshold are truly free of pulse pileup effects. \citet{zhang2024observation} investigate this with a scaled, counts-based comparison of GECAM-C and the GBM BGO over the 1-4~MeV energy range, finding reasonable agreement, suggesting negligible pulse pileup in the BGO data. However, the comparison neglects possible spectral evolution, and is not necessarily conclusive.

With the assumption that the BGO pulse pileup in these intervals is not significant, we can compare the GBM BGO-only spectral fits to the time intervals in \citet{zhang2024observation} to determine if we can verify their results. This is because the curvature in the continuum model and the line is contained within the BGO energy range. We perform spectral fits using both BGOs above 400~keV. In the 246-256~s interval our results show a large reduction in $\Delta$Pstat when including the line, whose centroid is comparable to the 37~MeV value in \citet{zhang2024observation}, but our measured width is significantly broader. Because the line centroid is near the GBM BGO high energy limit it is difficult to measure the line behavior. Additionally, while there is strong evidence for an additional component at these energies, because we cannot measure the behavior of the component at energies above the peak, we cannot be certain that it is a line. We are unable to constrain a line in the 270-275~s interval using only GBM BGO data. We do recover the line in the 275-280~s interval with similar centroids and widths as \citet{zhang2024observation}. Our $\Delta$Pstat value is $\sim$13, which appears significant given the results of our simulations for other intervals.

Thus, our results add further credence to the line beginning at least a few seconds before the 280~s start time in \citet{ravasio2024mega}. We note that this limitation was set by the public range for bad time intervals provided by the GBM Team, which was inclusive of all detector issues and not set on detector-specific intervals. Predicated on the discovery and temporal evolution of the line as measured in \citet{ravasio2024mega} it seems reasonable to use the parameter values measured from the earlier intervals identified in \citet{zhang2024observation} for characterization of the line behavior. However, there is less confidence in the line existence from GBM data before T0+275~s.

Continuing the prior discussions, there are, unfortunately, no observations which may confirm the line at times before those considered in \citet{ravasio2024mega}. The line description in \citet{zhang2024observation} is entirely contained only within the GBM BGO data. The Fermi Large Area Telescope Low Energy data is unusable due to saturation effects, and Astro‐Rivelatore Gamma a Immagini Leggero (AGILE) was not observing at the relevant time. No other facilities with coverage observe at these energies. 

\section{Conclusions}
The Ravasio Line achieves statistical discovery significance in multiple time intervals, confirming the major claims in \citet{ravasio2024mega}. There is no known instrumental effect which could have caused this spectral feature. Predicated on these two statements, the use of the earlier line intervals in \citet{zhang2024observation} is reasonable for characterization of the line, though these intervals are more ambiguous on the significance of the line.

\section*{Acknowledgments}
The USRA coauthors gratefully acknowledge NASA funding through cooperative agreement 80NSSC24M0035. E. Burns acknowledges many fruitful discussions with M. Negro.

\bibliography{bibliography}

\begin{thebibliography}{}
\expandafter\ifx\csname natexlab\endcsname\relax\def\natexlab#1{#1}\fi
\providecommand{\url}[1]{\href{#1}{#1}}

\bibitem[{Ackermann {et~al.}(2012)Ackermann, Ajello, Allafort, Atwood, Baldini, Barbiellini, Bastieri, Bechtol, Bellazzini, Bhat, {et~al.}}]{ackermann2012fermi}
Ackermann, M., Ajello, M., Allafort, A., {et~al.} 2012, The Astrophysical Journal, 745, 144

\bibitem[{Akaike(1974)}]{akaike1974new}
Akaike, H. 1974, IEEE transactions on automatic control, 19, 716

\bibitem[{Algeri {et~al.}(2019)Algeri, Aalbers, Mor{\aa}, \& Conrad}]{algeri2019searching}
Algeri, S., Aalbers, J., Mor{\aa}, K.~D., \& Conrad, J. 2019, arXiv preprint arXiv:1911.10237

\bibitem[{{Band} {et~al.}(1993){Band}, {Matteson}, {Ford}, {Schaefer}, {Palmer}, {Teegarden}, {Cline}, {Briggs}, {Paciesas}, {Pendleton}, {Fishman}, {Kouveliotou}, {Meegan}, {Wilson}, \& {Lestrade}}]{1993ApJ...413..281B}
{Band}, D., {Matteson}, J., {Ford}, L., {et~al.} 1993, \apj, 413, 281

\bibitem[{{Band} {et~al.}(1997){Band}, {Ford}, {Matteson}, {Briggs}, {Paciesas}, {Pendleton}, \& {Preece}}]{1997ApJ...485..747B}
{Band}, D.~L., {Ford}, L.~A., {Matteson}, J.~L., {et~al.} 1997, \apj, 485, 747

\bibitem[{{Band} {et~al.}(1994){Band}, {Ford}, {Matteson}, {Briggs}, {Paciesas}, {Pendleton}, {Preece}, {Palmer}, {Teegarden}, \& {Schaefer}}]{1994ApJ...434..560B}
---. 1994, \apj, 434, 560

\bibitem[{{Band} {et~al.}(1995){Band}, {Ford}, {Matteson}, {Briggs}, {Paciesas}, {Pendleton}, {Preece}, {Palmer}, {Teegarden}, \& {Schaefer}}]{1995ApJ...447..289B}
---. 1995, \apj, 447, 289

\bibitem[{{Band} {et~al.}(1996){Band}, {Ryder}, {Ford}, {Matteson}, {Palmer}, {Teegarden}, {Briggs}, {Paciesas}, {Pendleton}, \& {Preece}}]{1996ApJ...458..746B}
{Band}, D.~L., {Ryder}, S., {Ford}, L.~A., {et~al.} 1996, \apj, 458, 746

\bibitem[{Bhat {et~al.}(2014)Bhat, Fishman, Briggs, Connaughton, Meegan, Paciesas, Wilson-Hodge, \& Xiong}]{bhat2014fermi}
Bhat, P., Fishman, G., Briggs, M., {et~al.} 2014, Experimental Astronomy, 38, 331

\bibitem[{Bissaldi {et~al.}(2009)Bissaldi, von Kienlin, Lichti, Steinle, Bhat, Briggs, Fishman, Hoover, Kippen, Krumrey, {et~al.}}]{bissaldi2009ground}
Bissaldi, E., von Kienlin, A., Lichti, G., {et~al.} 2009, Experimental Astronomy, 24, 47

\bibitem[{Bowers {et~al.}(2017)Bowers, Smith, Martinez-McKinney, Kamogawa, Cummer, Dwyer, Wang, Stock, \& Kawasaki}]{bowers2017gamma}
Bowers, G.~S., Smith, D.~M., Martinez-McKinney, G., {et~al.} 2017, Geophysical Research Letters, 44, 10

\bibitem[{{Briggs} {et~al.}(1999){Briggs}, {Band}, {Preece}, {Paciesas}, \& {Pendleton}}]{1999ApL&C..39..237B}
{Briggs}, M.~S., {Band}, D.~L., {Preece}, R.~D., {Paciesas}, W.~S., \& {Pendleton}, G.~N. 1999, Astrophysical Letters and Communications, 39, 237

\bibitem[{{Briggs} {et~al.}(1996){Briggs}, {Band}, {Preece}, {Pendleton}, {Paciesas}, {Ford}, \& {Matteson}}]{1996AIPC..384..153B}
{Briggs}, M.~S., {Band}, D.~L., {Preece}, R.~D., {et~al.} 1996, in American Institute of Physics Conference Series, Vol. 384, Gamma-ray Bursts: 3rd Huntsville Symposium, ed. C.~{Kouveliotou}, M.~F. {Briggs}, \& G.~J. {Fishman} (AIP), 153--157

\bibitem[{Chaplin {et~al.}(2013)Chaplin, Bhat, Briggs, \& Connaughton}]{chaplin2013analytical}
Chaplin, V., Bhat, N., Briggs, M.~S., \& Connaughton, V. 2013, Nuclear Instruments and Methods in Physics Research Section A: Accelerators, Spectrometers, Detectors and Associated Equipment, 717, 21

\bibitem[{Chernoff(1954)}]{chernoff1954distribution}
Chernoff, H. 1954, The Annals of Mathematical Statistics, 573

\bibitem[{Gao \& Han(2012)}]{gao2012implementing}
Gao, F., \& Han, L. 2012, Computational Optimization and Applications, 51, 259

\bibitem[{Gross \& Vitells(2010)}]{gross2010trial}
Gross, E., \& Vitells, O. 2010, The European Physical Journal C, 70, 525

\bibitem[{Gruber {et~al.}(2014)Gruber, Goldstein, von Ahlefeld, Bhat, Bissaldi, Briggs, Byrne, Cleveland, Connaughton, Diehl, {et~al.}}]{gruber2014fermi}
Gruber, D., Goldstein, A., von Ahlefeld, V.~W., {et~al.} 2014, The Astrophysical Journal Supplement Series, 211, 12

\bibitem[{Jiang {et~al.}(2024)Jiang, Wei, Wang, Li, He, Ren, Wei, \& Jin}]{jiang2024discovery}
Jiang, L.-Y., Wei, D., Wang, Y., {et~al.} 2024

\bibitem[{Kaneko {et~al.}(2006)Kaneko, Preece, Briggs, Paciesas, Meegan, \& Band}]{kaneko2006complete}
Kaneko, Y., Preece, R.~D., Briggs, M.~S., {et~al.} 2006, The Astrophysical Journal Supplement Series, 166, 298

\bibitem[{Lesage {et~al.}(2023)Lesage, Veres, Briggs, Goldstein, Kocevski, Burns, Wilson-Hodge, Bhat, Huppenkothen, Fryer, {et~al.}}]{lesage2023fermi}
Lesage, S., Veres, P., Briggs, M., {et~al.} 2023, The Astrophysical Journal Letters, 952, L42

\bibitem[{{Meegan} {et~al.}(2009{\natexlab{a}}){Meegan}, {Lichti}, {Bhat}, {Bissaldi}, {Briggs}, {Connaughton}, {Diehl}, {Fishman}, {Greiner}, {Hoover}, {van der Horst}, {von Kienlin}, {Kippen}, {Kouveliotou}, {McBreen}, {Paciesas}, {Preece}, {Steinle}, {Wallace}, {Wilson}, \& {Wilson-Hodge}}]{Meegan2009}
{Meegan}, C., {Lichti}, G., {Bhat}, P.~N., {et~al.} 2009{\natexlab{a}}, \apj, 702, 791

\bibitem[{{Meegan} {et~al.}(2009{\natexlab{b}}){Meegan}, {Lichti}, {Bhat}, {Bissaldi}, {Briggs}, {Connaughton}, {Diehl}, {Fishman}, {Greiner}, {Hoover}, {van der Horst}, {von Kienlin}, {Kippen}, {Kouveliotou}, {McBreen}, {Paciesas}, {Preece}, {Steinle}, {Wallace}, {Wilson}, \& {Wilson-Hodge}}]{Meegan09}
---. 2009{\natexlab{b}}, \apj, 702, 791

\bibitem[{Mitchell {et~al.}(2022)Mitchell, Allman, \& Rhodes}]{mitchell2022generalized}
Mitchell, J.~D., Allman, E.~S., \& Rhodes, J.~A. 2022, arXiv preprint arXiv:2211.04136

\bibitem[{{Nemiroff}(1995)}]{Nemiroff95greatdebategrb}
{Nemiroff}, R.~J. 1995, \pasp, 107, 1131

\bibitem[{{Palmer} {et~al.}(1994){Palmer}, {Teegarden}, {Schaefer}, {Cline}, {Band}, {Ford}, {Matteson}, {Paciesas}, {Pendleton}, {Briggs}, {Preece}, {Fishman}, {Meegan}, {Wilson}, \& {Lestrade}}]{1994ApJ...433L..77P}
{Palmer}, D.~M., {Teegarden}, B.~J., {Schaefer}, B.~E., {et~al.} 1994, \apjl, 433, L77

\bibitem[{{Poolakkil} {et~al.}(2021){Poolakkil}, {Preece}, {Fletcher}, {Goldstein}, {Bhat}, {Bissaldi}, {Briggs}, {Burns}, {Cleveland}, {Giles}, {Hui}, {Kocevski}, {Lesage}, {Mailyan}, {Malacaria}, {Paciesas}, {Roberts}, {Veres}, {von Kienlin}, \& {Wilson-Hodge}}]{Poolakkil+21GBMspeccat}
{Poolakkil}, S., {Preece}, R., {Fletcher}, C., {et~al.} 2021, \apj, 913, 60

\bibitem[{Protassov {et~al.}(2002)Protassov, Van~Dyk, Connors, Kashyap, \& Siemiginowska}]{protassov2002statistics}
Protassov, R., Van~Dyk, D.~A., Connors, A., Kashyap, V.~L., \& Siemiginowska, A. 2002, The Astrophysical Journal, 571, 545

\bibitem[{Ravasio {et~al.}(2024)Ravasio, Salafia, Oganesyan, Mei, Ghirlanda, Ascenzi, Banerjee, Macera, Branchesi, Jonker, {et~al.}}]{ravasio2024mega}
Ravasio, M.~E., Salafia, O.~S., Oganesyan, G., {et~al.} 2024, Science, 385, 452

\bibitem[{Rutjes {et~al.}(2017)Rutjes, Diniz, Ferreira, \& Ebert}]{rutjes2017tgf}
Rutjes, C., Diniz, G., Ferreira, I., \& Ebert, U. 2017, Geophysical Research Letters, 44, 10

\bibitem[{Virtanen {et~al.}(2020)Virtanen, Gommers, Oliphant, Haberland, Reddy, Cournapeau, Burovski, Peterson, Weckesser, Bright, {et~al.}}]{virtanen2020scipy}
Virtanen, P., Gommers, R., Oliphant, T.~E., {et~al.} 2020, Nature methods, 17, 261

\bibitem[{Zhang {et~al.}(2024)Zhang, Xiong, Mao, Zhang, Xue, Zheng, Liu, Zhang, Wang, Ge, {et~al.}}]{zhang2024observation}
Zhang, Y.-Q., Xiong, S.-L., Mao, J.-R., {et~al.} 2024, Science China Physics, Mechanics \& Astronomy, 67, 1

\end{thebibliography}

\end{document}